\documentclass[sigconf, nonacm]{acmart}

\newcommand\vldbdoi{XX.XX/XXX.XX}
\newcommand\vldbpages{XXX-XXX}
\newcommand\vldbvolume{14}
\newcommand\vldbissue{1}
\newcommand\vldbyear{2020}
\newcommand\vldbauthors{\authors}
\newcommand\vldbtitle{\shorttitle} 
\newcommand\vldbavailabilityurl{URL_TO_YOUR_ARTIFACTS}
\newcommand\vldbpagestyle{plain} 

\usepackage{subcaption}
\begin{document}
\title{GTX: A Transactional Graph Data System For HTAP Workloads}

\author{Libin Zhou}
\affiliation{%
  \institution{Purdue University}
  \streetaddress{P.O. Box 1212}
  \city{West Lafayette}
  \state{Indiana}
  \postcode{47906}
}
\email{zhou822@purdue.edu}

\author{Walid G. Aref}
\orcid{0000-0002-1825-0097}
\affiliation{%
  \institution{Purdue University}
  \streetaddress{1 Th{\o}rv{\"a}ld Circle}
  \city{West Lafayette}
  \country{Indiana}
}
\email{aref@purdue.edu}

\begin{abstract}
Processing, managing, and analyzing dynamic graphs are the cornerstone in multiple application domains including fraud detection, recommendation system, graph neural network training, etc.
This demo presents GTX, a latch-free write-optimized transactional graph data system that supports high throughput read-write transactions while maintaining competitive graph analytics. GTX has a unique latch-free graph storage and a transaction and concurrency control protocol for dynamic power-law graphs. GTX leverages atomic operations to eliminate latches, proposes a delta-based multi-version storage, and designs a hybrid transaction commit protocol to reduce interference between concurrent operations. To further improve its throughput, we design a delta-chains index to support efficient edge lookups. GTX manages concurrency control at delta-chain level, and provides adaptive concurrency according to the workload. Real-world graph access and updates exhibit temporal localities and hotspots. Unlike other transactional graph systems that experience significant performance degradation, GTX is the only system that can adapt to temporal localities and hotspots in graph updates and maintain million-transactions-per-second throughput. GTX is prototyped as a graph library and is evaluated using a graph library evaluation tool using real and synthetic datasets.
\end{abstract}

\maketitle

\pagestyle{\vldbpagestyle}
\begingroup
\renewcommand\thefootnote{}\footnote{\noindent
This work is licensed under the Creative Commons BY-NC-ND 4.0 International License. Visit \url{https://creativecommons.org/licenses/by-nc-nd/4.0/} to view a copy of this license. For any use beyond those covered by this license, obtain permission by emailing \href{mailto:info@vldb.org}{info@vldb.org}. Copyright is held by the owner/author(s). Publication rights licensed to the VLDB Endowment. \\
\raggedright Proceedings of the VLDB Endowment, Vol. \vldbvolume, No. \vldbissue\ %
ISSN 2150-8097. \\
\href{https://doi.org/\vldbdoi}{doi:\vldbdoi} \\
}\addtocounter{footnote}{-1}\endgroup

\section{Introduction}
Dynamic graph management is an important part of many application domains including risk management, knowledge graph applications, recommendation services, etc~\cite{byte-graph-ref2,Sortledton}. Real-world graphs can reach hundreds of millions of vertices and billions of edges~\cite{Sortledton,konect-ref} while hundreds of thousands to millions of updates take place per second~\cite{byte-graph-ref2}. These applications need to run concurrent graph analytics while ingesting graph updates. Moreover, to concurrently run graph analytics and updates while preserving graph's consistency and avoiding anomalies, graph systems need to support read-write transactions~\cite{Teseo,chengmammoths,Sortledton}. Besides having high 
arrival
rate, graph updates also exhibit temporal localities and hotspots~\cite{Sortledton,konect-ref}. Namely, updates arriving at the same time frame likely belong to the same vertex (neighborhood) and a large number of such updates may arrive simultaneously. For example, at a certain time, lots of users are liking the same post on social media, creating a large amount of edges to the post simultaneously. In the rest of the paper, we  use "temporal localities" to collectively refer to graph updates with both temporal localities and hotspots. State-of-the-art transactional graph systems use coarse-grained concurrency control (e.g., vertex-centric locking~\cite{LiveGraph2,Sortledton}) 
and suffer from these update patterns~\cite{Sortledton}. To address these challenges and requirements, a graph system needs to support transactions, adapt to the temporal localities, and reduce the interference between concurrent transactions and analytics.

This demo presents GTX, a main memory latch-free transactional graph data system that is optimized for read-write transactions while maintaining competitive graph analytics. It targets the open problem of maintaining high throughput in workloads involving temporal localities~\cite{Sortledton}.
Many real world graphs are power-law graphs with hub vertices (vertices with large degree)~\cite{Sortledton,LiveGraph2}. For example, one of the real world graph edit-wiki has vertices with millions of adjacent edges and the edges' creation follows temporal localities and contains hotspots~\cite{Sortledton,konect-ref}. Most of the existing transactional graph systems only support transactions, but are unware of temporal localities. Moreover, they suffer from concurrently running updates and graph analytics. Two of the three state-of-the-art systems that we have evaluated have some bugs under such concurrent workloads. Therefore, we design GTX with the following features to resolve these challenges: 

\textbf{1.} GTX has a latch-free graph storage that uses atomic operations to update vertices and edges. It eliminates latching overheads and reduces thread idling.

\textbf{2.} GTX combines chain-based delta storage and linear delta storage to benefit from delta-chains' efficient lookup and linear storage's cache performance.

\textbf{3.} GTX has an efficient transaction management and concurrency control protocol by managing concurrency control at delta-chain level that adopts to the workload history.

\textbf{4.} GTX has a hybrid group commit protocol that improves transaction commit throughput. 

In this demo, we will present how GTX handles vertex and edge creation transactions of real world power-law graphs with and without temporal localities and how GTX executes read-write transactions while concurrently executing graphs analytics using synthetic graph logs in Graph Framework Evaluation(GFE)~\cite{Teseo,Sortledton}.
\section{Overview of GTX}
GTX addresses the problem 
of
managing and analyzing dynamic labeled property graphs. It consists of a latch-free adjacency list-based graph store and a transaction manager and a concurrency control protocol. GTX can manage both uniform and power-law graphs but is optimized for power-law graphs as many real-world graphs are power-law graphs~\cite{LiveGraph2}. In this demonstration, we focus on managing power-law graphs. 
\begin{figure*}[htbp] 
    \centering
    \includegraphics[width=\textwidth]{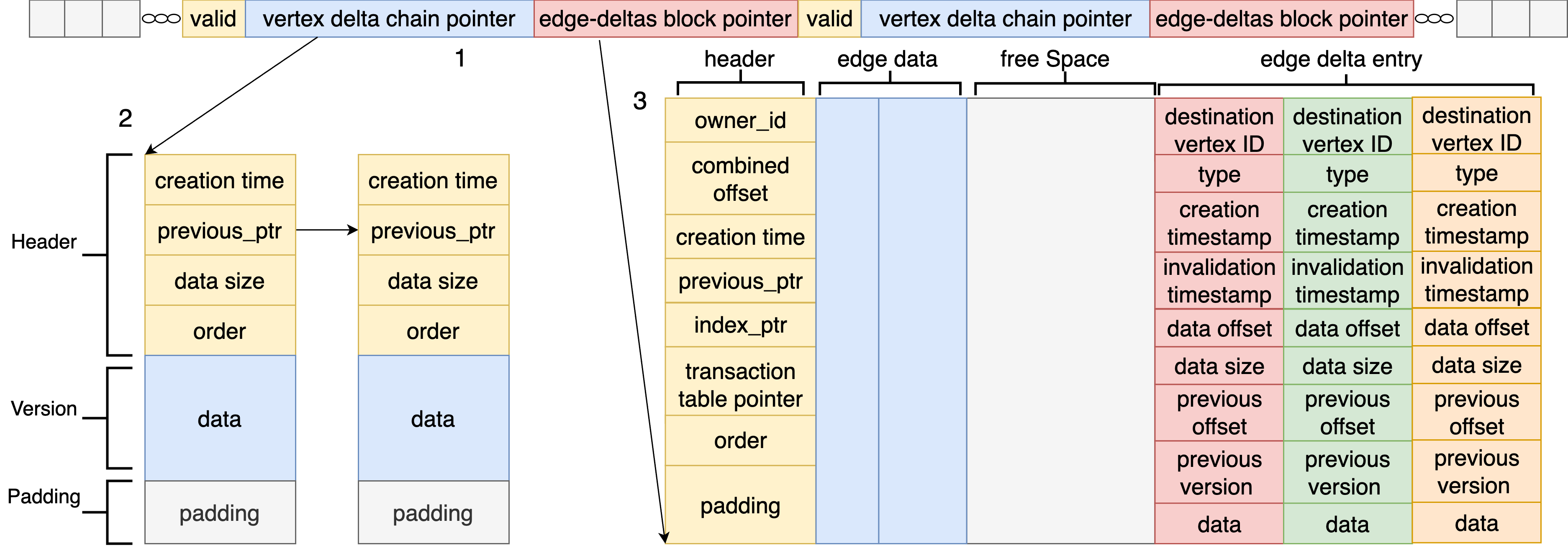}
    \caption{GTX Storage: 1.Vector-based vertex index (entry). 2.Vertex delta chain. 3.Edge-deltas block}
    \label{fig:storage}
\end{figure*}
GTX's graph store is a multi-version delta store, where each delta captures an operation (insert, update) of a vertex or  (insert, delete, update) of an edge. An overview of GTX's storage 
is given
in Figure~\ref{fig:storage}. Each edge-delta takes 1 cache line to avoid false sharing and is optionally associated with a variable-sized property that either resides inside the delta or at the other end of the block depending on the property size. GTX stores vertex versions as vertex delta-chains, and each vertex's adjacency list consecutively in a single edge-deltas block. Within each block, GTX organizes edge-deltas as delta-chains. Delta-chains support efficient single edge lookup and concurrency control and the linear edge-delta storage reduces random memory access, and improves cache performance during adjacency list scan. To support efficiently locating a vertex's 
delta-chain and edge-deltas block, GTX has a 
vertex index providing $O(1)$ lookup given vertex ID. Note that GTX supports edge labels, and stores edge-deltas (edges) of the same label in an edge-deltas block, but for simplicity of this demonstration, we assume all edges have the same label. GTX manages a delta-chains index of each edge-deltas block. Each edge-deltas block determines its delta-chains number at its allocation time, and allocates a vector where each vector entry $i$ stores the offset of the $i^{th}$ delta-chain within the block.

GTX adopts multi-version concurrency control~\cite{mvcc-ref}, and uses atomic operations to access and update vertex and edge versions, e.g. $fetch\_add$, $load$, and $compare\_and\_swap$($CAS$). Vertex versions are updated by a $CAS$ on the vertex delta-chains pointer. Each edge-deltas block maintains a {\em combined\_offset} that combines the offsets of the edge data region and edge-delta region (each 32 bits) into a 64 bits integer. Edge-delta allocation is done by executing a {\em fetch\_add} to the {\em combined\_offset} and write the edge-delta and its associated property data at the offsets returned by the operation.

GTX allocates a fixed number of worker threads that execute read-write transactions while additionally supports collectively running graph analytics using OpenMP~\cite{openmp-ref}. Besides a commit manager, GTX has no other server thread. Transactions' isolation and consistency ensures the correctness of concurrent operations. GTX read-write transactions support vertex inserts and updates, edge inserts, deletes, updates and lookups, and adjacency list scans. GTX guarantees atomicity of each transaction's updates. GTX implements all graph analytics under read-only transactions. While each read-write transaction is executed by its creator worker thread, a read-only transaction can be executed by several OpenMP threads concurrently. GTX further implements a state protection protocol to support edge-deltas block consolidation when it becomes overflow.

GTX adopts the block manager from~\cite{LiveGraph2} and manages garbage collection lazily. Memory blocks are recycled when no concurrent and future transactions can see them.

\section{GTX Transaction Operations}
Each GTX transaction is assigned a read timestamp ($rts$) at its creation time from a global read epoch, and it does not know its write timestamp ($wts$) until it gets committed. GTX guarantees Snapshot Isolation~\cite{SI-ref} of its transactions. Let $t$ refer to both the transaction and its transaction ID interchangeably. 
\subsection{Vertex Operations}
GTX transactions can insert new vertices and update them to new versions. Transaction $t$ reads a vertex $v$ by using an atomic $load$ on $v$'s delta-chains pointer after locating the $v$'s entry in the index. $t$ checks $v$'s delta-chain head, and compares its $rts$ against the vertex-delta's creation time. A transaction can see all deltas created before its $rts$. If the current vertex-delta represents the future version, $t$ uses the $previous\_ptr$ to read the previous vertex-delta (previous version) on the delta-chain. Vertex writes (insertion and updates) are handled by creating a new vertex-delta (new version), and invoking a $CAS$ on $v$'s delta-chain pointer. If the $CAS$ fails, $t$ observes write-write conflict and immediately abort. 
\subsection{Edge Write Operations}
GTX supports edge inserts, update and deletes, and they are handled similarly except having different delta types. GTX executes checked operations. An edge delete only takes place after determining the edge's existence. An edge insert only takes place after checking that the edge does not exist. Otherwise, the write operation is executed as an update.
$t$ inserts/updates/deletes an edge, say $e(u,v)$, by creating an insert/update/delete edge-delta of $e(u,v)$. Edge-deltas blocks manage concurrency at the delta-chain level. $t$ calculates the delta-chain $e(u,v)$ belongs to, say the $i^{th}$ delta-chain, and locks 
it
in the delta-chains index. After locking the delta-chain, $t$ searches the latest edge-delta(version) of $e(u,v)$ using each edge-delta's $previous\_offset$, and writes $t$ as its invalidation timestamp if the edge-delta exists. Then, $t$ allocates memory by updating the {\em combined\_offset}, and write the edge-delta in the corresponding locations. The edge-delta stores the offset of the current $i^{th}$ delta-chain head, and stores the offset of the previous edge-delta of $e(u,v)$. Moreover, it stores $t$ as the new edge-delta's creation timestamp. If $t$ fails to lock the $i^{th}$ delta-chain, $t$ will abort.
\subsection{Edge Read Operation}
GTX supports two types of edge read operations: Single-edge lookup and adjacency list scan. Single-edge lookup uses the delta-chains index and is identical as checking whether an edge exists in an edge write operation. Adjacency list scan is implemented as scanning the whole edge-deltas block. After locating the edge-deltas block, $t$ loads the {\em combined\_offset}. The {\em combined\_offset} determines where in the edge-deltas block the scan starts. $t$ scans all edge-deltas starting from there,
and can see all edge-deltas created before and invalidated after (or not invalidated at all) $t$'s $rts$.
\subsection{Hybrid Commit}
GTX supports a hybrid group commit protocol. GTX manages a transaction table that provides $O(1)$ access time of looking up a transaction's status. A transaction registers a commit request by informing the commit manager. The commit manager keeps a global write epoch number, and assigns its value as the $wts$ to the commit group of transactions by updating their status in the transaction table. After committing each group, the commit manager increments the global read epoch and global write epoch by 1. 
The committing transactions eagerly find and update their deltas' timestamps with their commit timestamp. Concurrent transactions may observe a delta's timestamp is a committed transaction (ID) by querying the transaction table, 
and update the delta's timestamp for the committing transaction. We call this \textbf{cooperative} commit protocol hybrid commit. It reduces group commit latency, and saves cost in synchronizing committing transactions and commit manager.
\subsection{Block Consolidation and Garbage Collection}
Edge-deltas blocks may become overflowed and require allocating a new block for the most recent edge-deltas while recycling the old block with the outdated edge-deltas. GTX uses a latch-free state protection protocol inspired by~\cite{EPVS} for each edge-deltas block that eliminates latches in shared memory and preserves cache coherence. 
Its core is analyzing the edge-deltas block, allocating a new block according to workload history, and migrating the latest version edge-deltas to the new block while concurrent transactions can still read the original edge-deltas block. After consolidating an edge-deltas block, 
the old deltas block is considered outdated and is placed in a queue. GTX tracks timestamps of current running transactions, and periodically has its worker threads free the memory blocks no transactions can access.
\subsection{Transaction Example}
\begin{figure*}[htbp] 
    \centering
    \includegraphics[width=\textwidth]{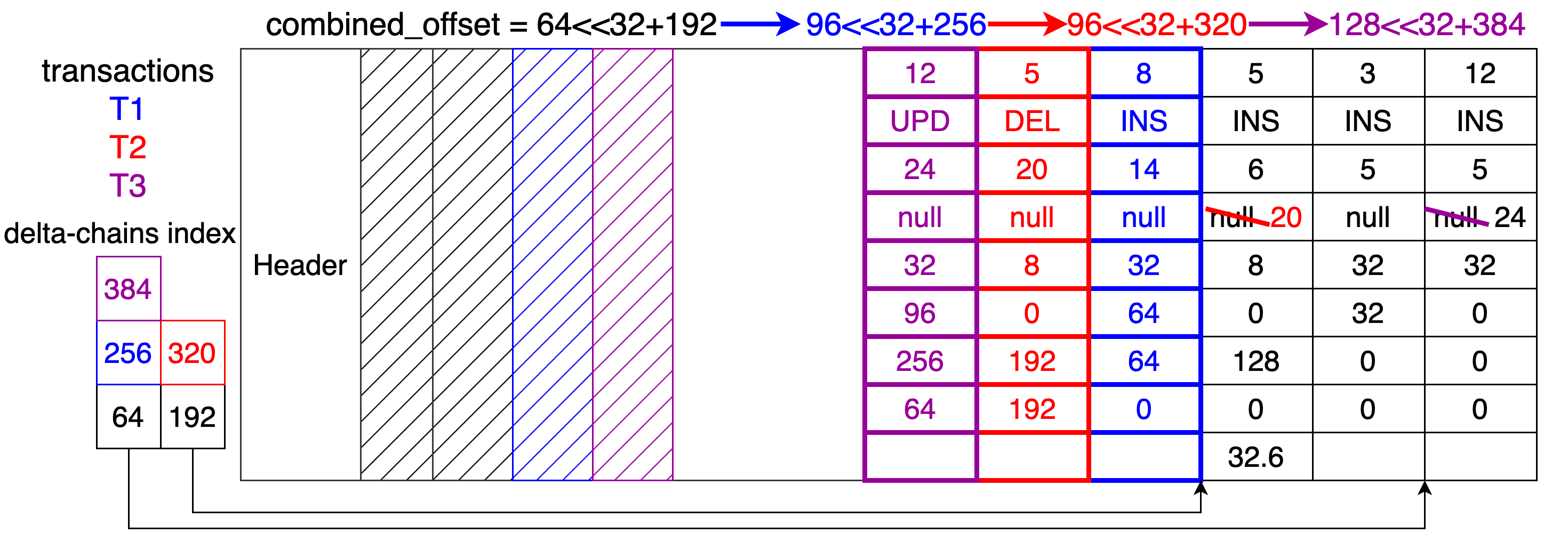}
    \caption{Examples}
    \label{fig:example}
\end{figure*}
We present how GTX transactions lookup and update edges. 
We use $ts$, $ts_c$, $ts_r$, $ts_{cr}$, and $ts_i$ to indicate {\em timestamp,  commit\_timestamp, read\_timestamp,  creation\_timestamp}, and {\em invalidation\_timestamp}, respectively.
Figure \ref{fig:example} illustrates how transactions execute edge insert, delete, and update. Transactions and their modification to the edge-deltas block and delta-chains index are in the same color. The edge-deltas block for Vertex 1 originally stores the 3 edges $(1,12)$, $(1,3)$, and $(1,5)$ using 3 edge-deltas and 2 delta-chains. Delta-chains Index Entry 0 stores $offset=64$ that points to $(1,12)$, and Entry 1 stores $offset=192$ pointing to $(1,5)$. Transaction $T_1$ inserts a new edge $(1,8)$ with 32 bytes of property data. $T_1$ calculates that Edge $(1,8)$ belongs to Delta-Chain $0=8 \bmod 2$ where 2 is the number of delta-chains, and uses the $0^{th}$ delta-chains index entry $64$ to locate the edge-deltas chain, and searches for a previous version of $(1,8)$. Then, it allocates space in both the delta region and the data region by updating {\em combined\_offset} atomically. Next, it writes the new insert edge-delta and its property into the allocated region. Note that the new edge-delta has {\em previous\_offset=64} and {\em previous\_version=0} because it needs to point to the previous edge-delta on the delta-chain, and Edge $(1,8)$ has no previous version. At commit phase, $T_1$ updates delta-chains index entry to point to the new edge-delta, and the commit manager commits $T_1$ at $ts_c =14$. The edge-delta's $ts_{cr}$ gets updated to 14. Next, Transaction $T_2$ deletes Edge $(1,5)$. $T_2$ uses the delta-chains index entry 1 to find the previous version of $(1,5)$ at offset $196$. Then, it allocates the space for its edge-delta by updating the {\em combined\_offset}. Since $T_2$ is deleting $(1,5)$, it does not allocate space in the data region. $T_2$ creates its delete edge-delta, and stores offset $192$ as the new delta's previous version. $T_2$ commits in a similar way by updating the delta-chains index Entry 1 and registers with the commit manager. After the commit manager commits $T_2$ at $ts_c=20$, the delete edge-delta's $ts_{cr}$ and the previous version edge-delta's $ts_i$ are both updated to $20$. Transaction $T_3$ updates Edge $(1, 12)$ with a new 32 bytes property. $T_3$ finds the previous version of $(1,12)$ at offset 64 using the delta-chain similarly. $T_3$ allocates space and creates an \textbf{update} edge-delta for $(1,12)$ and commits, and the delta-chains index, the new edge-delta's $ts_{cr}$, and the previous version's $ts_i$ get updated.
\vspace{-0.3 cm}
\begin{table}[htbp]
\begin{tabular}{ |c|c|c|c| } 
 \hline
 Dataset & Category & Vertex Count & Edge Count\\ 
 yahoo-songs & real-world & 1,625,951 & 256,804,235 \\ 
 edit-wiki & real-world & 50,757,442 & 572,591,272 \\ 
 graph500-24 & synthetic & 8,870,942 & 260,379,520 \\
 \hline
\end{tabular}
\caption{Dataset Statistics}
\label{fig:metadata}
\vspace{-1 cm}
\end{table}
\section{Demonstration and Evaluation}
We demonstrate GTX's abilities to ingest real-world graphs with hotspots and to efficiently execute concurrent read-write transactions and graph analytics. We evaluate GTX's performance against other state-of-the-art transactional graph systems, illustrating GTX's high throughput in transactions, and robustness across workloads while maintaining competitive graph analytics performance.
\subsection{Demonstration Scenario}
We showcase GTX's high transaction throughput for workloads with temporal localities, while maintaining competitive graph analytics performance. We compare GTX's transaction throughput and graph analytics latency against state-of-the-art transactional graph systems.
We adopt the Graph Framework4 Evaluation (GFE) from Teseo~\cite{Teseo} and Sortledton~\cite{Sortledton} for evaluation and demonstration. We use two real-world power-law graph datasets yahoo-songs(y-s) and edit-wiki(e-w)~\cite{konect-ref} and two sets of synthetic graph update logs (2.6 billion entries each) generated by~\cite{graphlog-ref} on graph500-24~\cite{graphalytics-paper2}. One graph log shuffles edge updates and the other contains temporal localities. The statistics of the datasets can be found in Table~\ref{fig:metadata}.

We execute the demonstration on a dual-socket machine with Intel(R) Xeon(R) Platinum 8368 CPU @ 2.40GHz processors with 156 CPU and 192GB of DRAM. We run all evaluations in a single NUMA node if the memory size fits. All systems execute checked operations as transactions, checking an edge's existence before performing the write. This demonstration only considers undirected graphs and each system treats an undirected edge as two directed edges. Therefore, for each edge $e(u,v)$, 
each
system creates a transaction that checks whether $e(u,v)$ and $e(v,u)$ exist, and inserts both edges.
\begin{table}[htbp]
    \vspace{-0.3 cm}
    \begin{tabular}{|c|c|c|c|c|}
     \hline
         &  y-s shuffled & y-s ordered & e-w shuffled & e-w ordered\\
         GTX & 6727920 & 4864010 & 3909976 & 3946243\\
         Sortledton & 4118089 & 443473 & 3686403 & 432638\\
         Teseo & 3534455 & 105234 & 2609071 & 24640 \\
         LiveGraph & 723299 & 151309 & 603040 & 102398\\
      \hline
    \end{tabular}
    \caption{Graph Construction Throughput(txns/s)}
    \label{tab:construction}
    \vspace{-1 cm}
\end{table}
\begin{table}[htbp]
    \begin{tabular}{|c|c|c|c|c|}
     \hline
           & PR & SSSP & PR hotspot & SSSP hotspot\\
         GTX  & 4365903 & 4265851 & 4959926 & 4265851 \\
         Sortledton & 3568179 & 3490729 & 1364880 & N/A\\
         LiveGraph & 640138 & 630250 & 418388 & 405232\\
     \hline
    \end{tabular}
    \caption{Mixed-Workload Throughput(txns/s)}
    \label{tab:mixed-throughput}
    \vspace{-1 cm}
\end{table}

\begin{table}[htbp]
    \begin{tabular}{|c|c|c|c|c|}
     \hline
          & PR & SSSP & PR hotspot & SSSP hotspot\\
         GTX  & 15993070 & 22575448 & 14020837 & 18414454 \\
         Sortledton & 10525814 & 15560585 & 10391524 & N/A\\
         LiveGraph & 13720317 & 21072157 & 10633277 & 18571507\\
     \hline
    \end{tabular}
    \caption{Mixed-Workload Latency(microsecond)}
    \label{tab:mixed-latency}
    \vspace{-1 cm}
\end{table}

\subsection{Evaluation}
We run two sets of experiments to show GTX's high performance. The first experiment constructs real-world power-law graphs with (ordered) and without (shuffled) temporal localities to evaluate each system's transaction throughput (shown in Table~\ref{tab:construction}). Then, we make each system process update logs with and without temporal localities and concurrently run graph analytics (pagerank(PR) and single source shortest path(SSSP)). We report the transaction throughput and graph analytics latency in Table~\ref{tab:mixed-throughput},~\ref{tab:mixed-latency}.
The results demonstrate significant advancement in performance of GTX over other systems w.r.t. transcation throughput and analytics workloads. 

\begin{acks}
Walid G. Aref acknowledges the support of the National Science Foundation under Grant Number IIS-1910216.
\end{acks}

\bibliographystyle{ACM-Reference-Format}
\bibliography{sample}

\end{document}